# High-pressure high-temperature solution growth, structural, and superconducting properties of Fe-substituted $MgB_2$ single crystals

N.D. Zhigadlo [a*], R. Puzniak [b]

[a] *CrystMat Company, CH-8037 Zurich, Switzerland*

[b] *Institute of Physics, Polish Academy of Sciences, Aleja Lotnikow 32/46, PL-02668 Warsaw, Poland*

**Abstract**

Clarifying the impact of Fe doping on the structural and superconducting properties of $MgB_2$ is crucial, considering that iron is commonly used as a sheath material for the fabrication of metal-clad $MgB_2$ wires and tapes. To date the effects of Fe doping have only been investigated in polycrystalline samples, but the obtained results are controversial. Here, we report the successful growth of $Mg_{1-x}Fe_xB_2$ single crystals in a quaternary Mg-Fe-B-N system using the cubic anvil high-pressure and high-temperature technique. The reaction took place in a closed boron nitride crucible at a pressure of 3 GPa and a temperature of 1960 °C. The grown crystals exhibit plate-like shapes with sizes up to $0.9 \times 0.7 \times 0.1$ $mm^3$. The variation of the critical temperature $T_c$ of $Mg_{1-x}Fe_xB_2$ crystals with Fe content was found to be different from that observed in polycrystalline samples. For a small Fe doping, up to $x \leq 0.03$, the behaviour of $T_c(x)$ is similar to that for the crystals with Al and C substitutions, which suggests that Fe is in non-magnetic state. In this doping range, measurements of the temperature-dependent magnetization performed in high magnetic fields exclude spin states other than $S = 0$ for the Fe ions. However, for $x \geq 0.03$, certain crystals start to show a dramatic decrease in $T_c$, suggesting that Fe might be in a magnetic state. The *M-H* dependence for these crystals shows significant increase of magnetization with increasing field in low magnetic field, pointing to a weak ferromagnetism. Overall, the availability of Fe substituted $MgB_2$ single crystals exhibiting such peculiar behaviour offers a unique opportunity to investigate the effect of disorder alone on one hand and the influence of magnetic substituent on the superconducting characteristics on the other.

## 1. Introduction

Magnesium diboride ($MgB_2$) has drawn a considerable attention among the scientists since, despite being a simple intermetallic compound, it exhibits a remarkably high transition temperature, $T_c$, of about 39 K [1]. As most diborides it belongs to the *P6/mmm* space group and possesses a simple hexagonal structure. Magnesium diboride consists of honeycomb boron layers separated by magnesium layers. Each Mg atom is located at the center of hexagons of B atoms. The hexagonal boron layers are similar to graphite, where hybrid $sp^2$ orbitals form in the planes thus leading to a strong covalent bonding. In addition, the boron $p_z$ orbitals along the *c* axis create a π bond similar to graphite.

The intense theoretical and experimental research has led to a consensus that $MgB_2$ is a multiband, namely σ and π, conventional *s*-wave phonon mediated BCS-type superconductor [2-5]. Each band has its own energy gap, which is reduced as the temperature approaches the critical $T_c$ value [6]. Various techniques, including point-contact spectroscopy [7], tunnelling spectroscopy [8], Raman spectroscopy [9,10], and photoemission spectroscopy [11,12] not only confirm the presence of two superconducting gaps, but they also provide an estimate for them. The energy gap of the σ bands goes from 6.4 to 7.2 meV (with an average value of 6.8 meV), while for the π bands, it ranges from 1.2 to 3.7 meV (with an average value of 1.8 meV) [7-12]. It is established that the covalent 2D σ-band is of hole type and is mainly responsible for the superconductivity in $MgB_2$, whereas the metallic 3D π-band is of electronic type with a comparatively negligible contribution (see, Fig. 1) [2-6].

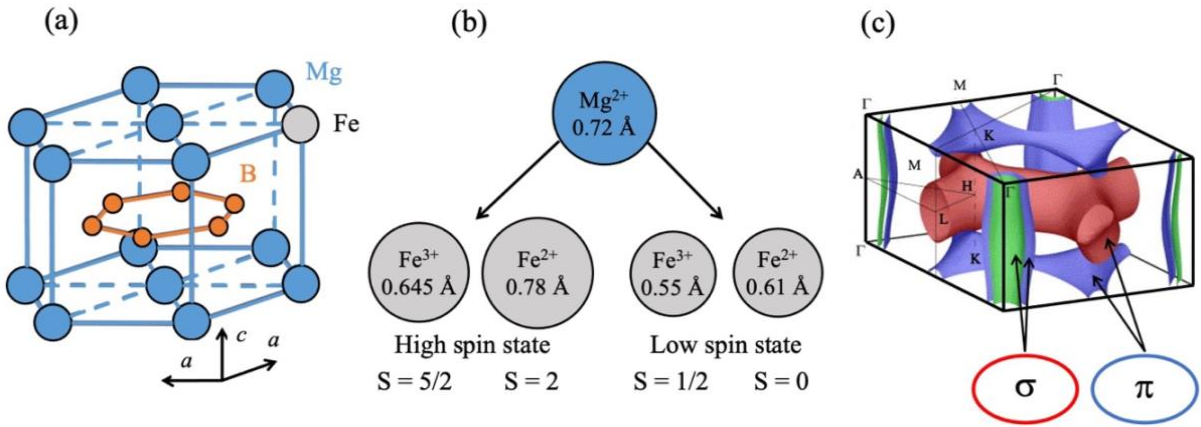

**Fig. 1.** Crystal structure of $MgB_2$ (a), possible spin state of Fe ions (b) with the values of ionic radii given according to Ref. 13, Fermi surface of $MgB_2$ (c). The Fermi surface of $MgB_2$ consists of two σ sheets along the Γ-Γ lines and two tubular structures defining a honeycomb lattice and arising from the π states.

In spite of the fact that $MgB_2$ is one of the simplest binary compounds, some of its physical, chemical, and electrical properties are very intriguing and not yet fully understood even after numerous investigations. For example, the idea of high-temperature topological superconductivity in $MgB_2$ still requires further experimental verification [12,14,15].

Beyond the purely scientific interest, the $MgB_2$ superconductor has attracted increasing attention from the scientific community for its potential applications. Compared to the high-temperature superconductors, such as YBa$_2$Cu$_3$O$_7$ and (Bi,Pb)$_2$Sr$_2$Ca$_2$Cu$_3$O$_{10}$ and conventional superconductors like Nb-Ti and Nb$_3$Sn, it offers a number of advantages [16]. Magnesium diboride has a comparatively high $T_c$, higher than those of conventional metallic superconductors, which enables the closed-cycled cooling of MgB$_2$ at 20 K using a suitable cryocooler. In contrast to other high-temperature superconductors, MgB$_2$ has lower anisotropy, larger coherence length, and better current flow across the grain boundaries and an inexpensive growth process that does not require rare-earth elements. Additionally, crystal orientation is not required to produce MgB$_2$ wires, contrary to (Bi,Pb)$_2$Sr$_2$Ca$_2$Cu$_3$O$_{10}$ and YBa$_2$Cu$_3$O$_7$, where uniaxial and biaxial orientation is needed, respectively.

Many efforts have been devoted to the study of the chemical substitutions of MgB$_2$, focusing on understanding the nature of the parent compound as well as on the possible improvements for technological applications [16-20]. However, until now, only the substitution of Al for Mg and C for B has been possible at relatively high doping levels [21,22]. Other substitutions for Mg$_{1-x}$M$_x$B$_2$ such as M = Ti, Zr, Mo, Mn, Fe, Ca, Ag, Cu, Y, La, Dy, Ho, Nd, etc., are restricted to a very low concentration [5,16,20,23-26]. The alkaline earth metals (Ca, Sr, and Ba) do not substitute on the Mg site, but form hexaborides (CaB$_6$, SrB$_6$, and BaB$_6$). Another possibility is that a possible ternary phase may prove to be more stable than the doped MgB$_2$ phase, so that little or no doping occurs.

Among the attempted transition-metal substitutions, that with Fe is rather unique and it is also of particular importance as a practical cladding metal or as a diffusion barrier for MgB$_2$ wire fabrication [27-28]. Typically, for practical applications, it is necessary to prepare MgB$_2$ thin films on metal substrates, which in turn enables the fabrication of MgB$_2$ superconducting wires. Common metals like Fe, Cu, and Al are examples of candidate metal substrates. Among these, Al is the most unstable because its post-annealing temperature of 650 °C is close to its melting point. Cooper is also unsuitable since the post-annealing decreases the $T_c$ and critical current density, $J_c$, significantly due to an interaction between MgB$_2$ and Cu. Therefore, currently Fe is the most promising metal substrate. Although Fe was not detected in the MgB$_2$ thin films fabricated on Fe tape, it is hypothesized that during annealing a tiny quantity from the Fe substrate diffuses through the B layer into the MgB$_2$ layer, potentially affecting the superconductivity of the material [29].

The effects of Fe doping on the superconducting and structural properties of MgB$_2$ have been studied so far on polycrystalline samples only and there is still uncertainty as to how much Fe can be introduced into MgB$_2$ [30-33]. It is worth mentioning that a considerably high solubility, up to $x = 0.4$, has been reported for Mg$_{1-x}$Fe$_x$B$_2$ [34,35]. However, since the lattice parameters and the $T_c$ of the doped compound do not change systematically with Fe concentration, one may speculate that iron enters only partially in the MgB$_2$ crystal structure, with a substantial amount of impurity phases formed in these samples. In polycrystalline samples, in addition to the MgO impurity, the Fe-based FeB and FeB$_2$ impurities are usually formed. Consequently, the Fe concentration which was taken at nominal value could be highly uncertain and overestimated. In this regard, it is important to examine single crystalline samples to understand which level of Fe substitution is accessible in MgB$_2$ and whether Fe ions carry a magnetic moment in $Mg_{1-x}Fe_xB_2$. The hexagonal symmetry of $MgB_2$ implies that many of its properties might be anisotropic. Therefore, single-crystal studies are necessary to elucidate this matter. It is evident that a homogeneous Fe doping of $MgB_2$ is still a significant challenge, which is not easy to achieve by low-temperature synthesis methods, but it may be overcome by using high temperatures. In this work, we present the results of high-pressure high-temperature solution growth of Fe-doped $MgB_2$ single crystals and report on their structural and superconducting properties.

2. Experimental details

Single crystals of $Mg_{1-x}Fe_xB_2$ were grown under high pressure using a cubic anvil press. Magnesium flakes (Fluka, 99.99% purity), iron powders (Alfa Aesar, >99.99%), amorphous boron (Alfa Aesar, >99.99%), and amorphous boron nitride (Saint-Gobain Advanced Ceramics, >99%) powders were used as starting materials. Amorphous boron and boron nitride were annealed under dynamic vacuum at 1200 °C for 1 h to minimize the contamination of oxygen and other impurities. Starting materials with various nominal contents were well mixed and pressed into a pellet. The pellet was placed in a BN crucible of 8 mm inner diameter and 9 mm long. The heating element was a graphite tube. Six tungsten carbide anvils generated pressure on the whole assembly. The temperature of the sample was determined by the pre-calibrated relationship between the applied electrical power and the measured temperature in the cell. A water-cooling system prevented the overheating of the anvils. More details regarding the high-pressure apparatus and the experimental setup can be found elsewhere [36,37]. We have previously used the high-pressure method with great success to grow crystals of a variety of superconducting [38-40] and magnetic materials [41-43], diamonds [44], cuprate oxides [45,46], pyrochlores [47,48], polymers [49], and a number of other compounds [50-53].

The phase purity of the obtained crystals was confirmed by x-ray diffraction. The Fe content and elemental mapping of the produced crystals were determined by energy dispersive x-ray (EDX) and transmission electron microscopy (TEM) spectroscopies. For all Fe substitutions from 0.3% to 3.6%, the crystals are single phase and homogenous, at least within ± 0.03% of Fe content. The lattice parameters of Fe-substituted crystals were determined at room temperature using a four-circle single-crystal x-ray diffractometer Siemens P4 with Mo $K_{\alpha 1}$ radiation, $\lambda = 0.71073$ Å. The same set of 32 reflections recorded in the $15° < 2\theta < 32°$ range was used to calculate the unit cell parameters. The detailed study of the structure was carried out on a Bruker SMART diffractometer equipped with a charge-coupled device (CCD) detector with Mo $K_{\alpha 1}$ radiation. The refinement of $Mg_{1-x}Fe_xB_2$ structure with Fe on Mg position was successful and no phase separation was observed.

Magnetic properties in the normal and superconducting states were investigated by magnetic moment measurements performed as a function of temperature and field with a Quantum Design magnetic property measurement system (QD-MPMS) equipped with a 7 T superconducting magnet. In order to determine the upper critical field, the magnetic moment $m$ was measured at constant field upon heating from the zero-field-cooled state (ZFC mode) or the field-cooled state (FC mode), with a temperature sweep of 0.1 K/min. Occasionally, $m$ was also measured at constant temperature with increasing field, using the step-by-step option. Complementary torque measurements were performed to obtain the upper critical field properties. The torque $\tau = m \times B = m \times H$ was recorded as a function of the angle between the applied field and the $c$-axis of the crystal for various fixed temperatures and fields. For the torque measurements, a QD physical property measurement (QD-PPMS) with torque option and a maximum field of 9 T was used [54].

3. Results and discussion

Conventional crystal growth methods do not work for pure $MgB_2$ or substituted $MgB_2$ analogs. High temperature solution growth in metals such as Al, Cu, etc., at normal pressure, as used for other borides, is not possible due to very low $MgB_2$ solubility in such metals or formation of other compounds [55]. At ambient pressure, the solubility of $MgB_2$ in Mg is extremely low up to the Mg boiling temperature (1107 °C). At higher temperatures, the solubility increases, but the partial pressure of Mg vapor above molten Mg also increases and it reaches about 100 bar at 1960 °C. This means that for growing $MgB_2$ single crystals from a Mg solution, a high-temperature, high-pressure technique should be used [56,57]. Indeed, studies of the phase diagram have shown that, to stabilize $MgB_2$ at high temperatures, not only high-pressure Mg vapor, but also high hydrostatic pressure is necessary. At high pressure the gas phase disappears and only the solid and liquid phases remain in reciprocal equilibrium. This allows the high-temperature crystal growth of pure $MgB_2$ and of various substituted systems from solution [19,21,22,58,59]. Based on these previous successful achievements, we applied the same methodology to the growth of Fe substituted $MgB_2$ crystals.

To substitute iron in $MgB_2$ crystals, part of Mg in the precursor was replaced by Fe. Starting materials with various nominal Fe contents were tested. For example, Mg:Fe:B:BN in a 9.5:0.5:12:1 ratio resulted in crystals with 0.5% Fe substituted $MgB_2$. The main challenge when growing single crystals of $Mg_{1-x}Fe_xB_2$ is the relatively poor reactivity of Fe. This is probably the main reason why it is so difficult to synthesize single-phase polycrystalline samples at ambient conditions [31-35]. Using higher synthesis temperatures at high pressure increases iron solubility.

In a typical growth process, a pressure of 30 kbar was applied at room temperature and then in 1 h the temperature was increased up to the maximum value of 1960 °C, kept there for 1 h, and then decreased to room temperature within 1.5 h. Crystals of the $Mg_{1-x}Fe_xB_2$ phase grew from a solution in magnesium, where iron diffuses in the melt and partially replaces magnesium. After that the solidified lump was heat treated in a quartz ampoule at 700 °C to remove the excess of Mg. As a result, $Mg_{1-x}Fe_xB_2$ plate-like single crystals of sizes up to $0.9 \times 0.7 \times 0.1$ mm$^3$ could be grown (Fig. 2).

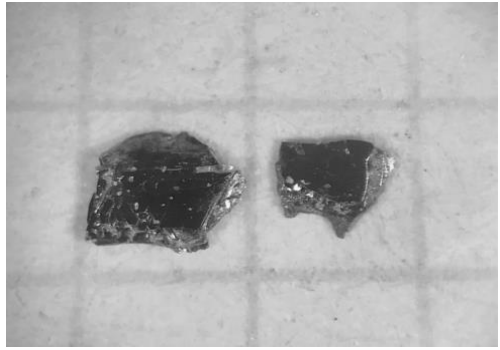

**Fig. 2.** Optical image of single crystals of (Mg,Fe)B$_2$ grown at high pressure and high temperature, here shown on a 1 mm grid paper.

As illustrated in Fig. 3(a,b), the lattice parameters were found to decrease with increasing Fe content, indicating that Mg is successfully replaced by Fe atoms. The variation of the *a*-axis parameter with *x* is much smaller than that in C-substituted crystals, but comparable with Al- and Mn-substituted crystals [19,21,22,58]. The behaviour of *c(x)* has the same tendency as for Al- and Mn-substituted crystals, i.e., it decreases with increasing Fe content, whereas for C-substituted crystals it remains almost constant.

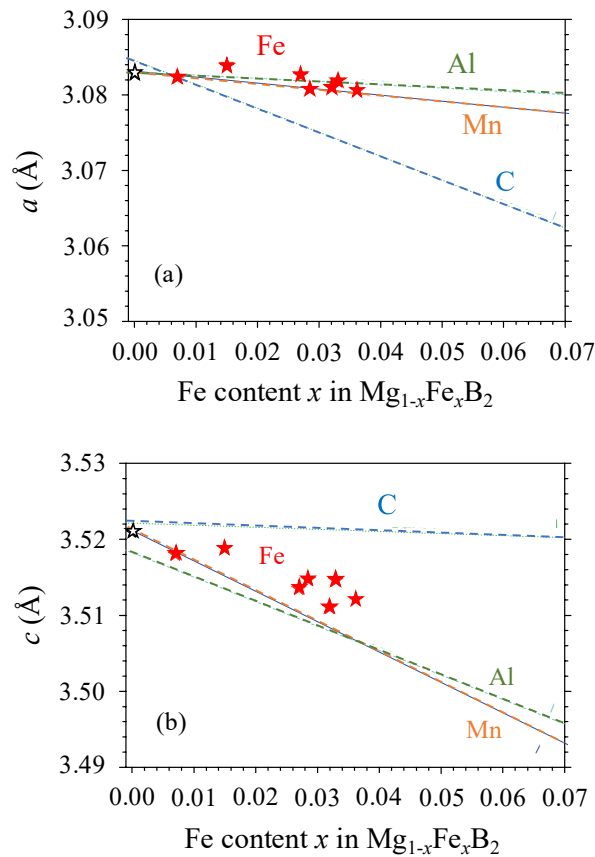

**Fig. 3.** (a,b) Lattice parameters *a* and *c* versus Fe content *x* (determined with EDX) for the series of Mg$_{1-x}$Fe$_x$B$_2$ single crystals. The dashed lines represent the lattice parameters for Al-, C-, and Mn-substituted crystals (Refs. [21], [22], and [58]).

Based on the ionic radii of Mg and Fe, it is natural to conclude that the distinct contraction of the MgB$_2$ unit cell reflects the fact that Fe atoms substitute on the Mg site rather than on the B site [13]. A similar conclusion was also derived from single-crystal x-ray investigations, where the Mg$_{1-x}$Fe$_x$B$_2$ structure was refined with Fe on the Mg position only. As Table 1 shows, the changes in structural parameters due to iron substitution are not very large. This indicates that substitution of Fe for Mg does not substantially alter the overall crystal structure.

| Fe content, $x$ | $a$, Å | $c$, Å | $c/a$ | $V$, Å$^3$ | $T_c$, K |
|---|---|---|---|---|---|
| 0.0 | 3.0834(2) | 3.5215(1) | 1.142(1) | 28.99(4) | 38.8 |
| 0.007 | 3.0823(8) | 3.5180(9) | 1.141(3) | 28.94(7) | 35.8 |
| 0.015 | 3.0840(7) | 3.5188(8) | 1.140(9) | 28.98(6) | 34.9 |
| 0.029 | 3.0808(9) | 3.5147(3) | 1.140(8) | 28.89(1) | 33.6 |
| 0.032 | 3.0815(1) | 3.5114(1) | 1.139(6) | 28.87(5) | 26.1 |
| 0.036 | 3.0818(9) | 3.5146(3) | 1.140(4) | 28.90(9) | 33.1 |

**Table 1.** The estimated iron content ($x$), lattice parameters ($a$, $c$), their ratios ($c/a$), lattice volumes ($V$), and superconducting temperatures ($T_c$) of selected Mg$_{1-x}$Fe$_x$B$_2$ single crystals. The lattice is hexagonal with a $P6/mmm$ space group and has one formula unit per cell.

A much weaker substitution effect on the lattice parameters was found in polycrystalline Mg$_{1-x}$Fe$_x$B$_2$, where the Fe content was assumed to coincide with its nominal value. Hence, one may speculate that Fe enters only partially in the crystal structure, while a substantial amount of impurity phases is formed in these samples [23]. As a consequence, the Fe concentration which was taken as the nominal one could be overestimated. Figure 4 shows TEM elemental mapping of 3.6% Fe-substituted MgB$_2$ crystals. As can be seen, the Mg, Fe, and B maps are uniformly distributed across the sample with no indication of agglomerates.

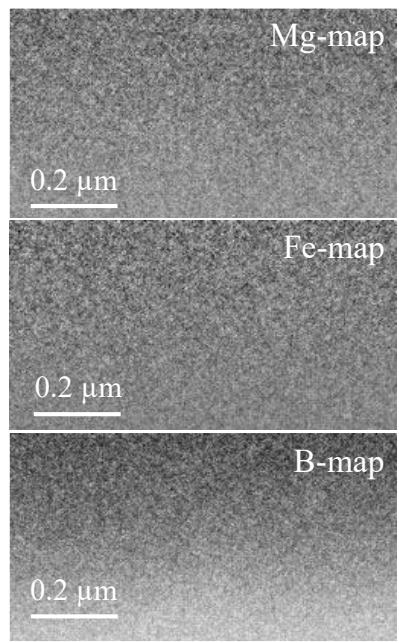

**Fig. 4.** TEM elemental mapping images of 3.6% Fe-substituted MgB$_2$ crystals. The Mg, Fe, and B containing particles are uniformly dispersed within the sample with no indication of agglomerates.

It is important to note that although the high-pressure route has produced doped single crystals, the doping levels reached at high pressure may not match the doping levels possible at low pressure due to the different thermodynamic conditions. In addition, the single crystals were grown at high temperature in a quaternary Mg-Fe-B-N system, which may have different phase boundaries than the ternary Mg-Fe-B system [56,57], i.e., the solubility limit of Fe in the Mg-Fe-B system may be different from the solubility limit of Fe in the Mg-Fe-B-N system. Also, the solubility may change as a function of temperature, which means that the dopant concentration may change with the cooling rate. All these factors need to be considered before attempting to use the results of doped single crystals to interpret those regarding the lower temperature/pressure samples.

Figure 5 shows the volume susceptibility versus temperature, illustrating the changes in $MgB_2$ superconducting transition as a result of the various Fe substitution rates. The data were taken under applied field of 3 mT on heating after cooling in zero field (ZFC) and then field cooling (FC). The transition temperature $T_c$ was defined by the intersection of the extrapolated susceptibility lines from the regions below and above the transition.

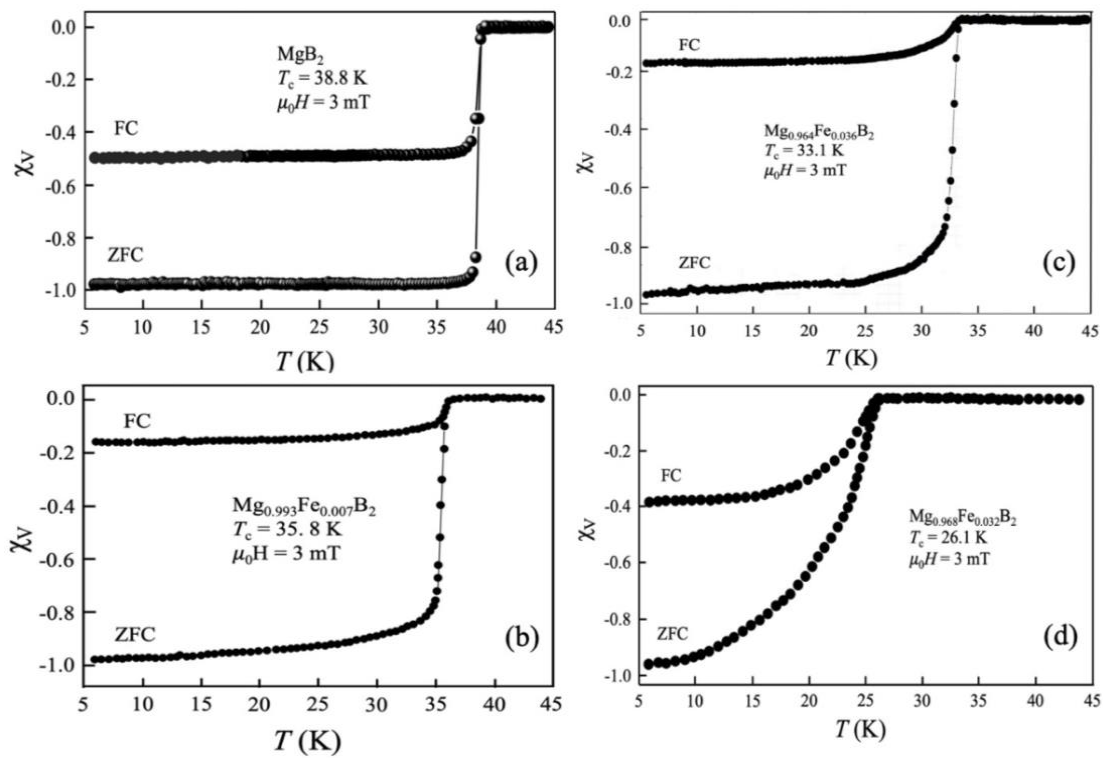

**Fig. 5.** Temperature dependence of volume susceptibility for $Mg_{1-x}Fe_xB_2$ single crystals containing different Fe content $x$: (a) = 0.0, (b) 0.007, (c) 0.036, and (d) 0.032).

All the samples show well-defined one-step transitions, as well as large shielding fractions indicating that the superconductivity is of bulk nature. The $T_c$ of the undoped crystal was 38.8 K, with a transition width of 0.5 K. For the Fe-substituted MgB$_2$ crystals, $T_c$ decreases upon increasing the doping level and the transition becomes broader. For example, at a 0.7% Fe doping level, $T_c$ drops by 3 K with a transition width of 2 K. The increased transition width with the increasing Fe content could be due to the disorder-induced scattering of the electrons. This correlates with the NMR results, where the addition of Fe in MgB$_2$ was shown to increase the inhomogeneity of the hyperfine field, thus resulting in a broader NMR spectrum. Such an effect usually broadens the band width and/or shifts the Fermi energy, leading to a change of the Fermi-level density of states (DOS). Surprisingly, NMR investigations on Fe substituted MgB$_2$ alloys have revealed that the density of states close to the Fermi level remains unaffected up to a substitution level of 3% [19]. Such finding is in strong contrast with the general behaviour observed for other types of substitution in MgB$_2$.

For most of the studied crystals $T_c$ decreases monotonously up to 3.6% of Fe substitution. However, already starting from 3% of Fe doping a strange behaviour is observed: some crystals start to show a rapid decrease in $T_c$. Such behaviour is unusual and has not been observed for any other substituents (Fig. 6) [19].

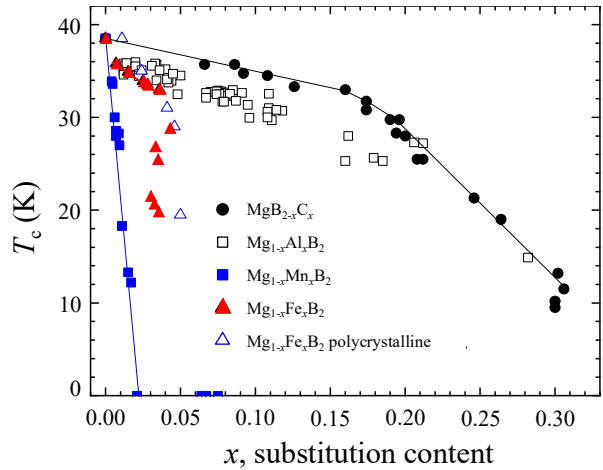

**Fig. 6.** Variation of $T_c$ for MgB$_2$ single crystals substituted with nonmagnetic (C, Al) and magnetic (Mn, Fe) ions. The main part of the results on C-, Al-, and Mn-substituted crystals has been published in Refs. [22], [21], and [58], respectively. Figure contains critical temperature $T_c$ as a function of Fe concentration for polycrystalline samples as well (as deduced from Ref. [35]).

The Fe substitution in MgB$_2$ may lead to a modification of its band structure, with two bands, σ and π, involved in superconductivity, therefore, influencing all the superconducting characteristics, in particular $T_c$, the upper critical field $H_{c2}$, and its anisotropy $\gamma_{Hc2}$ [60]. Compared with the impact of some non-magnetic substituents, such as Al and C [21,22], the effect of Fe doping on $T_c$ is slightly stronger, but not too drastic, which suggest that Fe is in a non-magnetic state. A quite different behaviour of $T_c$ was found in Mn-substituted MgB$_2$, where superconductivity was completely suppressed at 2% of Mn doping [58]. It was established that the key reason for the rapid $T_c$ drop in Mn-doped MgB$_2$ is the magnetic pair-breaking effect through spin-flip scattering [58,61,62]. Thus, one may speculate that the strong reduction in $T_c$ in some Fe-substituted MgB$_2$ crystals can be explained by the specific nature of Fe. For small Fe contents (up to $x \leq 0.03$) the behaviour of $T_c(x)$ is similar to that observed for Al substitution, which suggest that Fe is in a non-magnetic state. However, for $x \geq 0.03$ some crystals start to show a rapid decrease in $T_c$, implying that possibly Fe is in a magnetic state. Thus, we can conclude that substitution of $Fe^{2+}$ ions with a low-spin state ($S = 0$) results in a moderate reduction of $T_c$. However, at higher dopings, Fe may be in a high-spin state. For up to 3% Fe-substituted polycrystalline $MgB_2$ samples, a theoretical study and an ensuing NMR investigation showed the non-magnetic nature of Fe in the $MgB_2$ lattice [63].

The difference between high spin and low spin configurations of Fe ions lies in the way the electrons occupy the $d$-orbitals in response to the strength of the ligand field. High-spin configurations maximize the number of unpaired electrons, while low-spin configurations minimize the number of unpaired electrons by pairing them up in the low-energy orbits first. However, we note that the exact spin state of Fe in $Mg_{1-x}Fe_xB_2$ can be influenced by several factors, such as pressure, temperature, local crystal structure, the ligand field strength, the presence of other elements, and the geometry of the coordination environment, making it a complex interplay of conditions. At low levels of Fe substitution, Fe ions are likely to be in a non-magnetic low-spin state of $Fe^{2+}$, especially if they coordinate in a manner that favours strong ligand fields. At higher levels of Fe substitution, the situation may shift towards high-spin configurations. As Fe concentration increases, the local interactions may favour unpaired spins, particularly if weak-field ligands are present or if the environment becomes less effective at stabilizing low-spin configurations. High-spin Fe can lead to a magnetic behaviour due to the presence of unpaired electrons. In order to clarify this matter, we performed comparative magnetic measurements of Fe-substituted crystals with a moderate vs a sharp suppression of $T_c$.

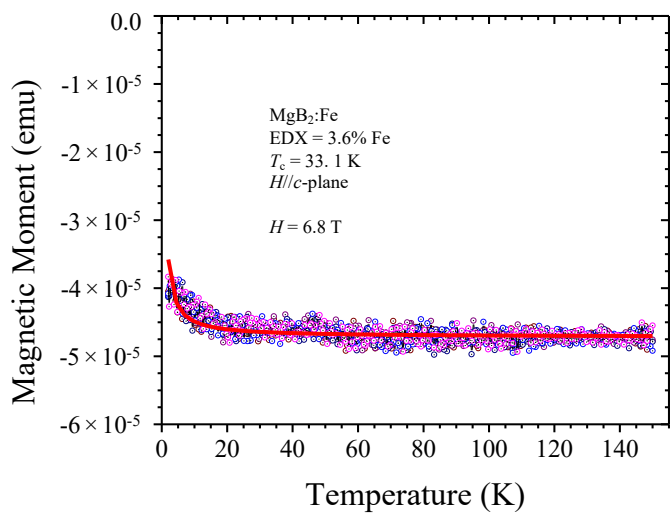

**Fig. 7.** Temperature dependence of the magnetic moment $m$ at a constant field $H = 6.8$ T for $MgB_2$ single crystal substituted with 3.6% of Fe. The measurements reveal Fe to be divalent in the low-spin configuration. $H$ was oriented parallel to the $c$-plane.

According to the temperature dependence of the magnetic moment $m$ at a constant field $H = 6.8$ T for MgB$_2$ single crystal substituted with 3.6% of Fe and $T_c$ of 33.1 K (see, Fig. 7), iron atoms are in a low-spin configuration of $Fe^{2+}$. Dominant contribution to magnetic moment recorded in full investigated temperature range represents temperature independent diamagnetism. There is small decrease of recorded signal in low temperature, by about 20% of full diamagnetic response, however, its temperature dependence is much weaker than that one given by simple addition of paramagnetic signal to temperature independent diamagnetism, as it is illustrated by red solid line in Fig. 7. Nuclear magnetic resonance spectroscopy studies and the Seebeck coefficient measurements of Fe-substituted MgB$_2$ alloys revealed that there is no magnetic moment associated with Fe dopants, and the density of states (DOS) close to the Fermi level is essentially unaffected by Fe substitution [63,64]. These previous observations, along with the current findings further suggest that the observed reduction in $T_c$ in Fe-substituted MgB$_2$ crystals cannot be explained by either the magnetic-induced pair breaking or by the reduction of the electronic DOS. We thus highlight the importance of the phonon contributions to the decrease of $T_c$, presumably attributed to the decreased phonon frequency of the $E_{2g}$ mode and/or the reduced electron-phonon coupling strength through Fe doping. In this regard, a further investigation using optical measurements would be instructive to verify which of these two factors is dominant.

The *M-H* dependence, measured at various temperatures from 10 to 150 K in a magnetic field ramping between -1 and 1 T, point to a weak ferromagnetism in the 3.2% of Fe-substituted crystal with sharp drop of $T_c$ at 26.1 K (Fig. 8). Sharp increase of magnetic signal with increasing magnetic field, on almost temperature independent diamagnetic background, is clearly visible in the low field and is almost temperature independent for the temperature above superconducting transition temperature. The jump of magnetization in low magnetic field at 10 K represents full diamagnetic response of bulk superconductor.

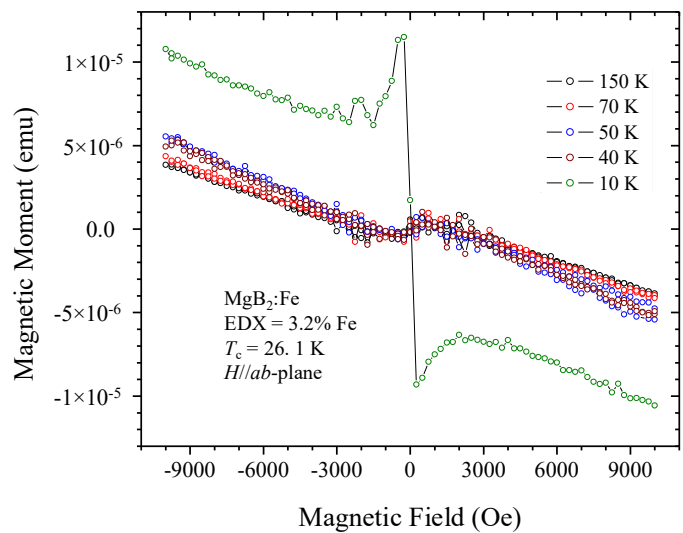

**Fig. 8.** Magnetic moment $m$ vs magnetic field for MgB$_2$ single crystal substituted with 3.2% of Fe. Magnetic field dependences were measured at temperatures 10-150 K in a magnetic field ramping between -1 and 1 T.

Observed field dependence of magnetization suggests that the 3.2% Fe-doped crystal exhibits a possibly magnetic state. To completely comprehend the magnetic properties of Fe ions in $MgB_2$, further measurements on a series of Fe substituted crystals are necessary.

Generally, attempts to substitute Mg for other elements had two main purposes: firstly, to find a way to increase $T_c$ in this system and secondly to introduce defects in the $MgB_2$ structure in order to decrease the mean free path of the normal electrons, decrease the coherence length, and consequently increase $H_{c2}$ and also to introduce local pinning centres to increase the irreversibility line [65-67]. To date, in all the substituted $MgB_2$ compounds, the critical temperature decreases upon increasing the content of the substituent atom. The partial Ca substitution for Mg in $MgB_2$ is expected to increase $T_c$ by decreasing the electronic bandwidth and, ultimately, by increasing the density of states at $E_F$, but no such substitution has been realized yet [68]. It was also claimed that $T_c$ can be further enhanced by applying a reasonable strain [69]. Various approaches have been attempted to enhance the superconducting properties of $MgB_2$, in particular $H_{c2}$, which is a key parameter for many superconducting applications [70-72]. The temperature dependence of the upper critical fields $H_{c2}^{//ab}$, $H_{c2}^{//c}$, and the upper critical field anisotropy $\gamma_{Hc2} = H_{c2}^{//ab}/H_{c2}^{//c}$ for the $MgB_2$ unsubstituted single crystals (diamonds; $T_c$ = 38.2 K) and Fe-substituted crystals (squares, triangles; 3.6% Fe, $T_c$ = 33.15 K) are shown in Fig. 9 (a,b). As can be seen in Fig. 9, the $H_{c2}$ values of Fe-substituted $MgB_2$ crystals are lower compared to pure $MgB_2$ crystals. The $\gamma_{Hc2}$ value is about 4.1 at 20 K (for $x$ = 3.6% Fe-substituted single crystal) and decreases to about 1.8 upon increasing the temperature.

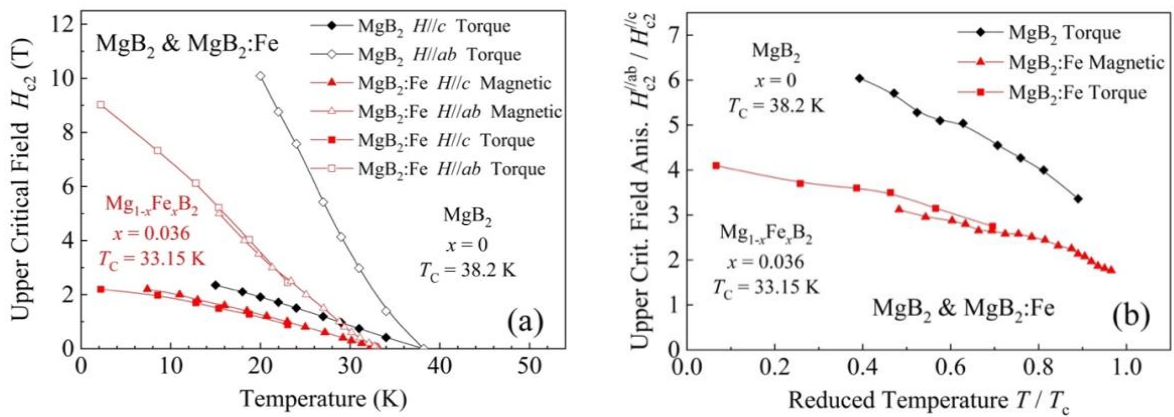

**Fig. 9.** Comparison of upper critical field (a) and its anisotropy (b), as determined via SQUID and torque magnetometry for $MgB_2$ and $Mg_{0.964}Fe_{0.036}B_2$.

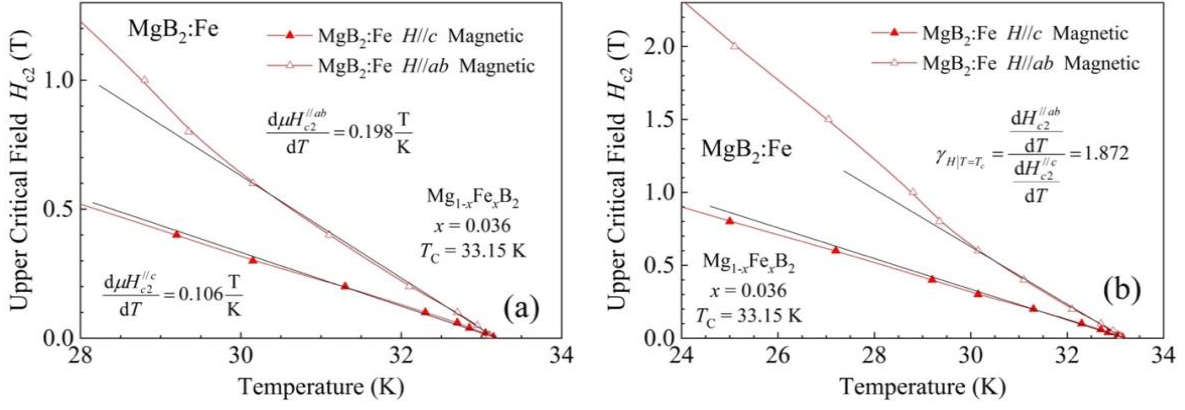

**Fig. 10.** Upper critical field (a) and its anisotropy (b) in the vicinity of $T_c$ for $Mg_{0.964}Fe_{0.036}B_2$ crystals determined with SQUID magnetometry.

For the $x = 0.036$ case, comparable results for the $H_{c2}$ value derived from magnetic measurements with the field aligned parallel and perpendicular to the $ab$ planes and with torque measurements were obtained. Two regions of the upper critical field curve are of particular interest, as they provide complementary information about the underlying electronic scattering. The slope near $T_c$ is given by the maximum of electron diffusivity in the two bands, while the limiting value of the upper critical field is determined by the minimum diffusivity. The diffusivities are different in the two bands, and they are anisotropic due to the 2D and 3D nature of the bands [69]. Therefore, special attention has been paid to obtain accurate $H_{c2}$ values at low fields to determine the upper critical field slope, $dH_{c2}/dT$, near $T_c$ (Fig. 10 (a)). For $H$ perpendicular the $ab$ plane, the $H_{c2}$ slope increases from ~ -0.1 T/K in unsubstituted crystals to ~ -0.106 T/K in 3.6% Fe-substituted $MgB_2$ crystals. In contrast, the upper critical field decreases upon Fe substitution when the field is parallel to the $ab$ plane, from ~ -0.22 T/K to ~ -0.198 T/K. Thus, the anisotropy near $T_c$ decreases when Fe is incorporated (Fig. 10 (b)). It is important to note that this reduction in anisotropy contrasts sharply with that seen in C-substituted crystals, where the $H_{c2}$ slope for both applied field orientations increase quickly upon C substitution, also leading to a reduction in anisotropy [22,73]. Hence, it is rather obvious that Fe substitution does not affect significantly the scattering in the electronic band which determines $H_{c2}$. This is further seen in the extrapolation of $H_{c2}$ toward zero temperature. Although the exact values of $H_{c2}$ are somewhat difficult to determine in $T$ sweeps, we observe that the $H_{c2}$ perpendicular to the $ab$ plane extrapolated to zero temperature is only slightly lower than that in the unsubstituted crystals (~3.5 T), thus suggesting the relevant minimum charge diffusivity to be unaffected by Fe substitution. For $H$ parallel to the $ab$ planes $H_{c2}(0)$ is reduced significantly. Some of these observations are qualitatively similar to previously reported [65,73].

## 4. Summary and conclusions

Using the cubic anvil high-pressure and high-temperature method, we were able to grow phase-pure Fe-substituted MgB$_2$ crystals in the quaternary Mg-Fe-B-N system up to a substitution level of 3.6%. Magnetic moment investigations and slight changes in $T_c$ for low Fe concentrations suggest that Fe ions in MgB$_2$ are in a low spin state ($S = 0$) and are non-magnetic up to at least a 3.6% substitution level. Nevertheless, for $x \geq 0.03$, certain crystals show a dramatic decrease in $T_c$, indicating that Fe might be in a magnetic state. The *M-H* dependence for these crystals show a somewhat strongly field dependent increase of magnetization in low magnetic field, pointing to a weak ferromagnetism. The availability of Fe-substituted MgB$_2$ single crystals with such peculiar behaviour offers the rare opportunity to investigate the impact of disorder on one hand and the influence of magnetic substituents on the superconducting characteristics on the other. Although the Fe doping investigated here was shown to have a negative effect on both $T_c$ and $H_{c2}$, this effect might not be as severe in Fe sheaths, since their synthesis temperature and the contact area between the sheath and MgB$_2$ are significantly smaller than in the case of Fe doped MgB$_2$ crystals.

The recently reported topological surface state in MgB$_2$ is expected to be highly tolerant against disorder and inadvertent doping changes. Therefore, the produced Mg$_{1-x}$Fe$_x$B$_2$ crystals may be a great platform for testing the high-temperature topological superconductivity hypothesis, in particular by using such powerful methods as high-energy resolution ARPES and scanning tunneling microscopy (STM).

**Acknowledgements**

The authors acknowledge support from the Laboratory for Solid State Physics, ETH Zurich where these studies were initiated. The authors also grateful to G. Schuck and S. Katrych for their help during the structural characterization and T. Shiroka for the critical reading of the manuscript and useful suggestions.